\documentclass[12pt,preprint]{aastex}
\usepackage{graphicx}
\usepackage{natbib}

\begin{document}

\title{The enormous outer Galaxy \ion{H}{2} region CTB~102}

\author{K. Arvidsson \and C. R. Kerton}
\affil{Department of Physics \& Astronomy, Iowa State University, Ames, IA, 50011; kima@iastate.edu, kerton@iastate.edu}

\and

\author{T. Foster}
\affil{Department of Physics \& Astronomy, Brandon University, 270-18th Street, Brandon , MB, R7A 6A9, Canada: fostert@brandonu.ca}

\date{Received; Accepted}

\begin{abstract}
We present new radio recombination line observations of the previously unstudied \ion{H}{2} region CTB~102. Line parameters are extracted and physical parameters describing the gas are calculated. We estimate the distance to CTB~102 to be 4.3 kpc. Through comparisons with \ion{H}{1} and 1.42 GHz radio continuum data, we estimate the size of CTB~102 to be $100-130$ pc, making it one of the largest \ion{H}{2} regions known, comparable to the W4 complex. A stellar wind blown bubble model is presented as the best explanation for the observed morphology, size and velocities.
\end{abstract}

\keywords{ISM: bubbles - ISM: \ion{H}{2} Regions - ISM: Individual (CTB~102) - Radio lines: ISM}

\section{Introduction}
The radio bright outer Galaxy region CTB~102 ($\ell=93 \fdg 060,~b=+2 \fdg 810$) was first cataloged by the \citet{ctb} radio survey of the Galactic plane. The source is then mentioned in subsequent Galactic radio surveys including \citet{kr80} where it is identified as KR~1. Using radio recombination line (RRL) observations at $\lambda$3~cm, \citet{lock89} (H87$\alpha$, $\sim$3$\arcmin$
beam) identified the region as a \ion{H}{2} region with a line brightness of $T_{line} = 18 \pm 1.3$~mK, a velocity of $V_{LSR} = -61.0 \pm 0.8$~km s$^{-1}$ and a full width at half maximum (FWHM) of $\Delta V =
 22.9 \pm 1.9$~km s$^{-1}$. Radio continuum images at 1.42 GHz and $\sim 1 \arcmin$ resolution from the Canadian Galactic Plane Survey \citep[CGPS,][]{tayl03}, show filamentary structures extending from a bright complex source. From the appearance of the structure and a kinematic distance estimate, the region appears to be a very large \ion{H}{2} region and a major feature in the Perseus spiral arm.

Yet this major Galactic region is unstudied. Suffering heavy extinction in this direction in the Galactic plane, there is no known optical counterpart to CTB~102.
The purpose of this study is to determine the basic properties of CTB~102, mainly how large in physical size it is, and how it influences its Galactic environment. In this paper we present new RRL observations towards CTB~102. RRL observations
allow direct velocity measurements, and along with
continuum observations will tell us the density and temperature of any gas in
the beam at or near thermodynamic equilibrium.

\section{Observations}
\begin{figure}
\plotone{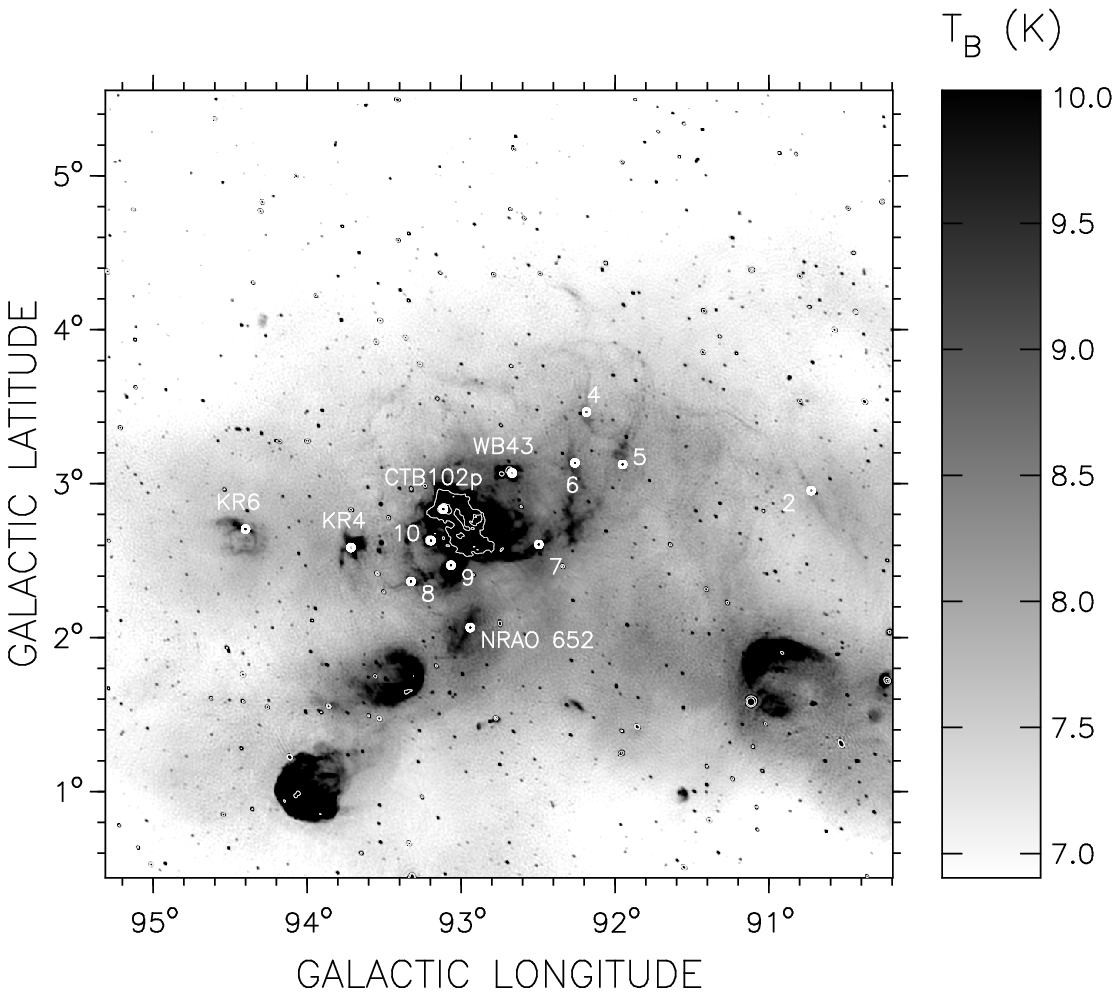}
\caption{CGPS 1.42 GHz continuum image with GBT pointings (see Table \ref{obs-tbl}) labelled and displayed as circles. The size of the circles correspond to the GBT beam size, $2.5 \arcmin$. The contours correspond to 15 and 30 K brightness temperature levels.\label{pointings}}
\end{figure}

\begin{figure}
\plotone{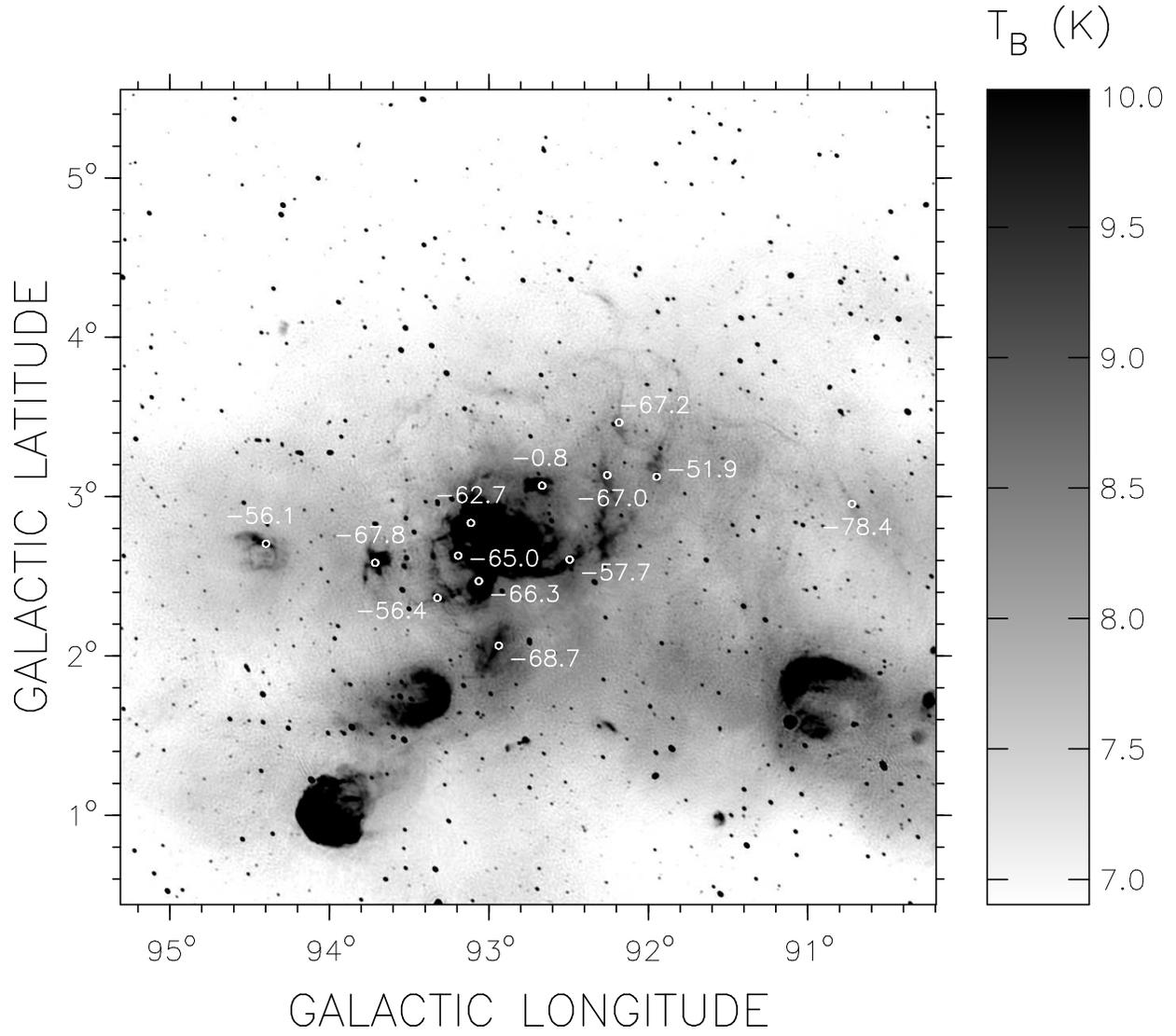}
\caption{As Figure \ref{pointings} but showing the observed velocities for each position.\label{velocities}}
\end{figure}

RRL observations towards CTB~102 were performed with the 100-m NRAO Green
Bank Telescope (GBT) during 6 nights in 2006, July 31, August 2-4, 15 \& 17. Twelve pointings were observed around the CTB~102 complex. These telescope pointings are illustrated in Figures~\ref{pointings} and \ref{velocities}. Positions and total integration times for the chosen observations are given in Table~\ref{obs-tbl}, the observations themselves were divided into 600~s scans. RRLs observed were H103$\alpha$ through H110$\alpha$. A 50 MHz bandwidth receiver was used to allow the eight recombination lines to be simultaneously observed in the high end of the $C$-band ($\nu = 4.8-5.9$ GHz). Both polarizations were admitted, and the spectrum consists of 4096 channels ($0.62-0.75$ km s$^{-1}$ per channel). System temperatures ranged from 19 to 26~K, depending mainly on the elevation
of the source. Average system temperatures for each observation are given in Table~\ref{obs-tbl}. As a check of the system's ability to record RRL emission, the bright {}``head'' of
CTB~102 at $\ell=93 \fdg 115,~b=+2 \fdg 835$, hereafter CTB102p, was observed for 600~s at the
beginning of each session.

\section{Data Reduction}
None of the  eight 50~MHz bands were seriously affected by radio frequency interference. Frequency-switched
scans in each linear polarization (YY, XX) were folded individually; since
frequency-switching was done in-band ($\pm$12.5~MHz), we doubled our effective
integration time. Each 600~s scan contains 16 spectra. To assure that no line structure (e.g. very extended wings from outflows) is removed, velocities forbidden by Galactic rotation (typically $- 300$~km s$^{-1} \leq V_{LSR} \leq -130$~km s$^{-1}$ and $+30$~km s$^{-1} \leq V_{LSR} \leq +200$~km s$^{-1}$) defines a range of baseline velocities. For each of the 16 spectra in every scan, a baseline was determined using a fourth-order polynomial fitted to the range of baseline velocities. This fitted baseline was then subtracted from every spectrum in each scan. After baseline subtraction, the spectra in the individual scans are combined (for every velocity channel, intensity values are summed up and then divided by the number of scans) to create averaged spectra, one for each line and polarization. At this point in the reduction process, there are 16 spectra (H103$\alpha$ XX, H103$\alpha$ YY, H104$\alpha$ XX, etc.) for every observation in Table~\ref{obs-tbl}.

These averaged spectra were regridded to a common channel width (0.67~km s$^{-1}$) and smoothed to a common velocity resolution (1.5015~km s$^{-1}$). Typically 4-10 of these averaged spectra do not show residual wavy baselines in regions of no RRL signal. The only exception is H110$\alpha$ (polarization XX), which shows a very wide ``bump'' in the spectrum centered around $-200$~km s$^{-1}$, extending into the region of the RRL signal. This line and polarization is completely excluded from the analysis. To reduce noise, composite
spectra, one for each filament, are made by combining the averaged spectra that do not show residual wavy baselines. The composite spectra typically have a noise level of $\Delta T_{rms} \la 1$~mK (antenna temperature).

Since RRLs are expected to be quite wide \citep[25-30~
km s$^{-1}$;][]{lock89}, a spectral resolution of 1.5~km s$^{-1}$ is unnecessarily fine. A
higher S/N can be achieved without loss of information by moderate smoothing of the composite spectra,
although too much will add an artificial width to spectral lines present. We
conservatively choose a resolution of 3.0~km s$^{-1}$.

\section{RRL Results and Analysis}
\begin{figure}
\plotone{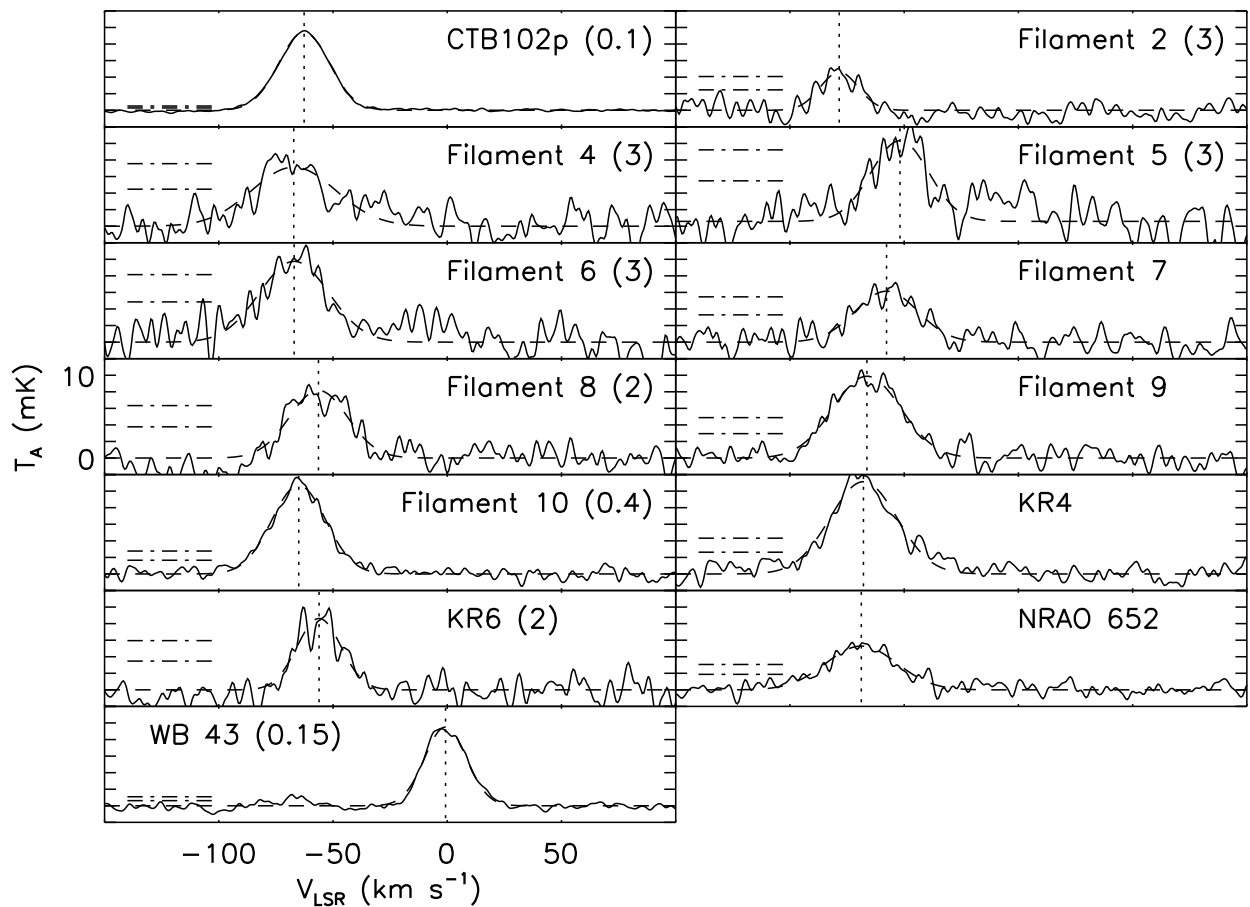}
\caption{Reduced spectra (solid) and Gaussian fits (dashed) for each observation, plotted on the same vertical and horizontal scale. The number in parenthesis is the scaling factor by which the spectrum was multiplied in order to fit the scale. The dash-dotted lines correspond to the $3 \sigma$ and $5 \sigma$ noise levels. The vertical dotted line is the central velocity from the Gaussian fit.\label{spectra}}
\end{figure}

To our final smoothed composite spectra, Gaussians are fit to obtain spectral line parameters: line amplitude ($T_l$), central velocity ($V_{LSR}$) and FWHM ($\Delta V$). Smoothed spectra and the Gaussian fits are shown in Figure~\ref{spectra} and the obtained parameters are presented in Table~\ref{par-tbl}. Note that antenna temperature has been divided by the beam efficiency, $\eta_b = 0.92$ for the GBT at 5 GHz, to convert to brightness temperature. The uncertainties in Table~\ref{par-tbl} are obtained in Monte Carlo fashion. To the originally obtained Gaussian fit, randomly drawn noise from a normal distribution with the same standard deviation as the previously obtained $\Delta T_{rms}$ is added. A new Gaussian is then fitted to the generated spectrum and its parameters stored. After 1000 repetitions, the standard deviation of the distributions of spectral line parameters is combined with the mean fit uncertainties to produce the total uncertainties. Figure~\ref{velocities} displays the observed velocities of the filaments/objects, overlayed on a CGPS 1.42 GHz continuum image.

\subsection{Velocities}\label{vel}
This study finds that the bright radio source CTB102p has a velocity of $V_{LSR} = -62.66 \pm0.05$ km~s$^{-1}$, which is 1.7 km~s$^{-1}$ less than the value in \citet{lock89}. The difference is probably due to the difference in pointing position, as well as the difference in telescope beam size between the studies. The velocity of CTB102p is hereafter referred to as $V_{ref}$. One purpose of this study is to find out the size of CTB~102, i.e. are the filaments and objects we see around CTB~102 at the same distance? The velocity gradient (for a flat rotation curve with $R_{\sun}=8.5$ kpc and $V_{\sun} = 220$ km~s$^{-1}$) in this part of the Galaxy is $\sim 10$ km s$^{-1}$ kpc$^{-1}$. Looking at the last column of Table~\ref{par-tbl}, most filaments and objects have a $|V-V_{ref}| \la 6$ km~s$^{-1}$ , so it seems likely that the CTB~102 complex and objects KR~4, KR~6 and NRAO~652 are in the same part of the Galaxy. In contrast, WB~43 ($-0.8$~km s$^{-1}$) is clearly a local \ion{H}{2} region, and in no way connected to the CTB~102 complex (Figure~\ref{spectra}).
Filament 2  is another possible unrelated object (see Sec~\ref{disc}), its velocity ($-78.4$~km s$^{-1}$) deviates the most from the velocity of CTB102p, $|V-V_{ref}| \sim 16$ km~s$^{-1}$, and is $\sim 10$ km s$^{-1}$ less than any other velocity observed in this study. Computing the average absolute difference for all observations (except WB~43) yields a result of 6.6 km~s$^{-1}$, but excluding filament 2 from the average, the value drops to 4.3 km~s$^{-1}$.

\subsection{Line Widths \& Electron Temperatures}\label{derived}
The majority of FWHMs in this study fall between 20~km s$^{-1}$ and 26~km s$^{-1}$, which is typical for dense \ion{H}{2} regions \citep{lock89}.

The line width of a RRL depends on mainly two things, thermal motion, which is described by the electron temperature ($T_e$), and microturbulence. Both of these processes produce Gaussian profile shapes, so a Doppler temperature ($T_D$), which provides an upper limit to the true electron temperature, can be adopted as follows \citep{tools}:
\begin{equation}
\Delta V = 0.21 T_D^{1/2}
\end{equation}
where $\Delta V$ is the observed FWHM in km~s$^{-1}$ from Table~\ref{par-tbl}, corrected for the spectral resolution of 3 km~s$^{-1}$.

The LTE electron temperature can be calculated using the line to continuum brightness ratio:
\begin{equation}
\frac{T_{l}}{T_c} \left( \frac{\Delta V}{\textrm{km}~\textrm{s}^{-1}} \right) = \frac{6.985 \times 10^3}{a(\nu,~T_e)} \left( \frac{\nu}{\textrm{GHz}}\right)^{1.1} \left( \frac{T_e}{\textrm{K}} \right)^{-1.15} \times \frac{1}{1+N_{\textrm{He}^+} / N_{\textrm{H}^+}}
\end{equation}
where $a(\nu,~T_{e})\simeq ~ 1$  and $N_{\textrm{He}^+}/N_{\textrm{H}^+}\simeq ~0.08$ \citep{tools}. The continuum brightness temperature $T_c$ is obtained by convolving the CGPS 1.42 GHz mosaic to the $2.5 \arcmin$ resolution of the GBT. The background is estimated individually for each filament (typical background estimate uncertainty $\sim 0.5$ K) and the background is subtracted from the average brightness temperature within a $2.5 \arcmin$ circle centered on the telescope pointing. Assuming temperature follows the power law $T_{\nu \arcmin} / T_{\nu} = \left( \nu / \nu \arcmin \right)^{-2.1}$ the resulting $T_c$ is scaled using the average frequency of the observed RRLs (5.46 GHz). The parameters derived in this way for each filament/object are in Table~\ref{der-tbl}. The $T_e$ is quite sensitive to the estimated background temperature, especially when the peak brightness temperature is barely above the background. Note that $T_D \geq T_e$ within the $1 \sigma$ uncertainty in all cases. $T_D$ provides a useful upper limit to $T_e$ in case the uncertainty in $T_e$ is large. These derived parameters for the ionized gas gives an average LTE electron temperature of $\sim 6 \times 10^3$ K. This value falls within the typical range of electron temperatures for \ion{H}{2} regions.

We note the FWHM of CTB102p determined in this study, $\Delta V = 18.34\pm0.08$~km s$^{-1}$, is 4.6~km s$^{-1}$ less than the value for $\Delta V$ in \citet{lock89}. Again, the difference is probably due to the difference in beam size and pointing position.

\section{Distance}
With the majority of the filaments' velocities found to be very similar to that of the central
body, the whole CTB~102 complex subtends about one to two degrees of sky. Its
distance would be able to tell us how far across its influence extends across
the outer disk. Here we provide the first ever distance estimate for this
\ion{H}{2} region.

\begin{figure}
\epsscale{0.5}
\plotone{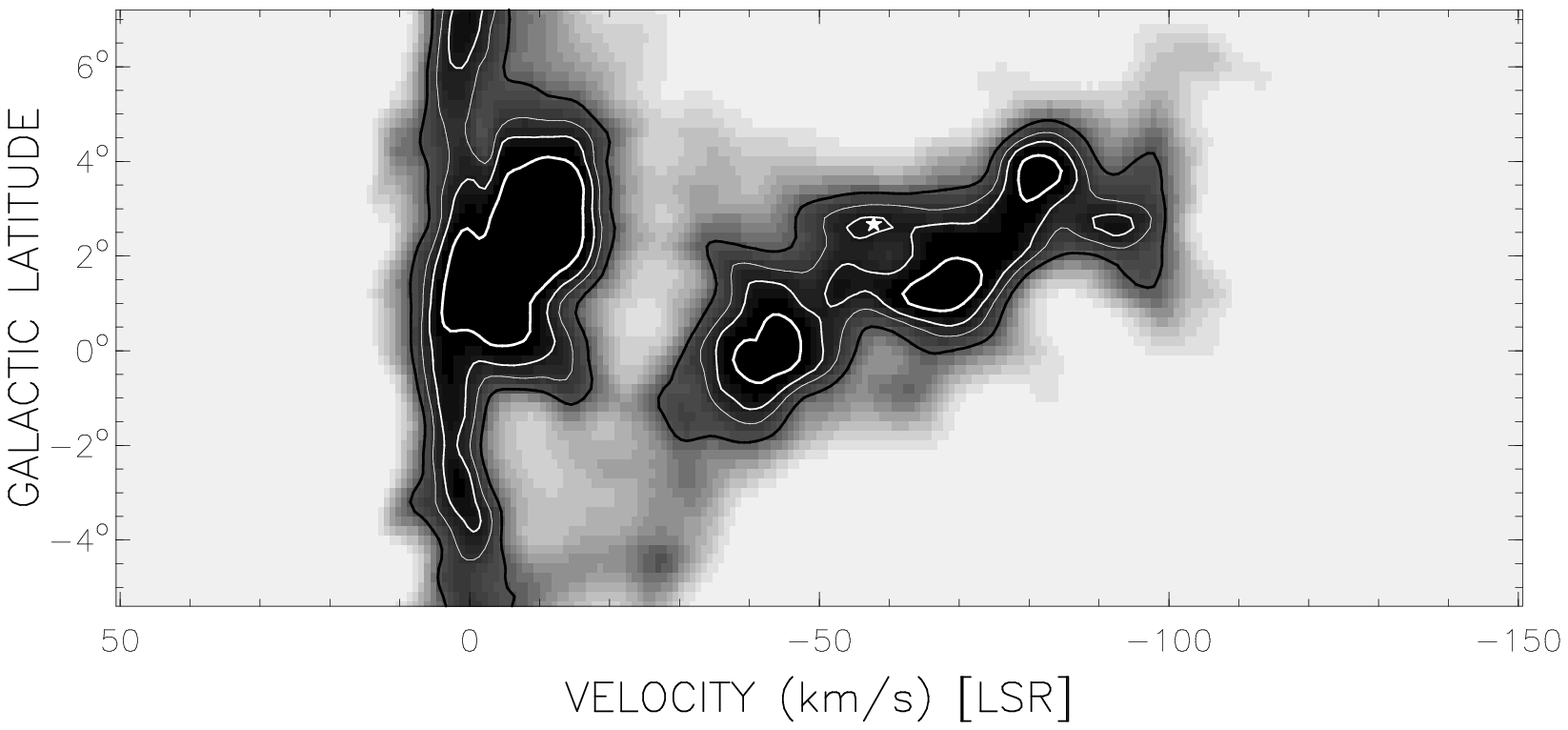}
\plotone{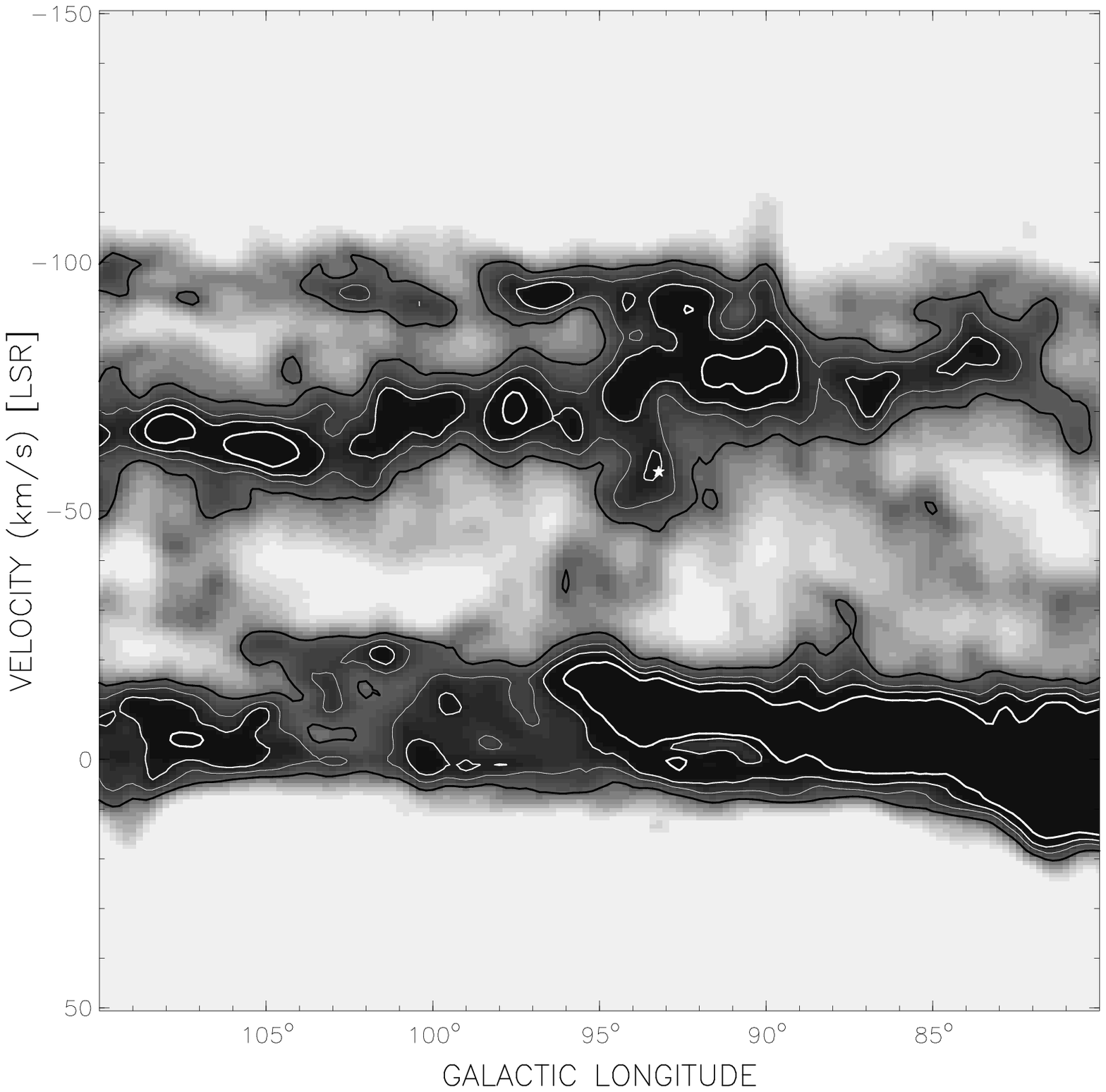}
\plotone{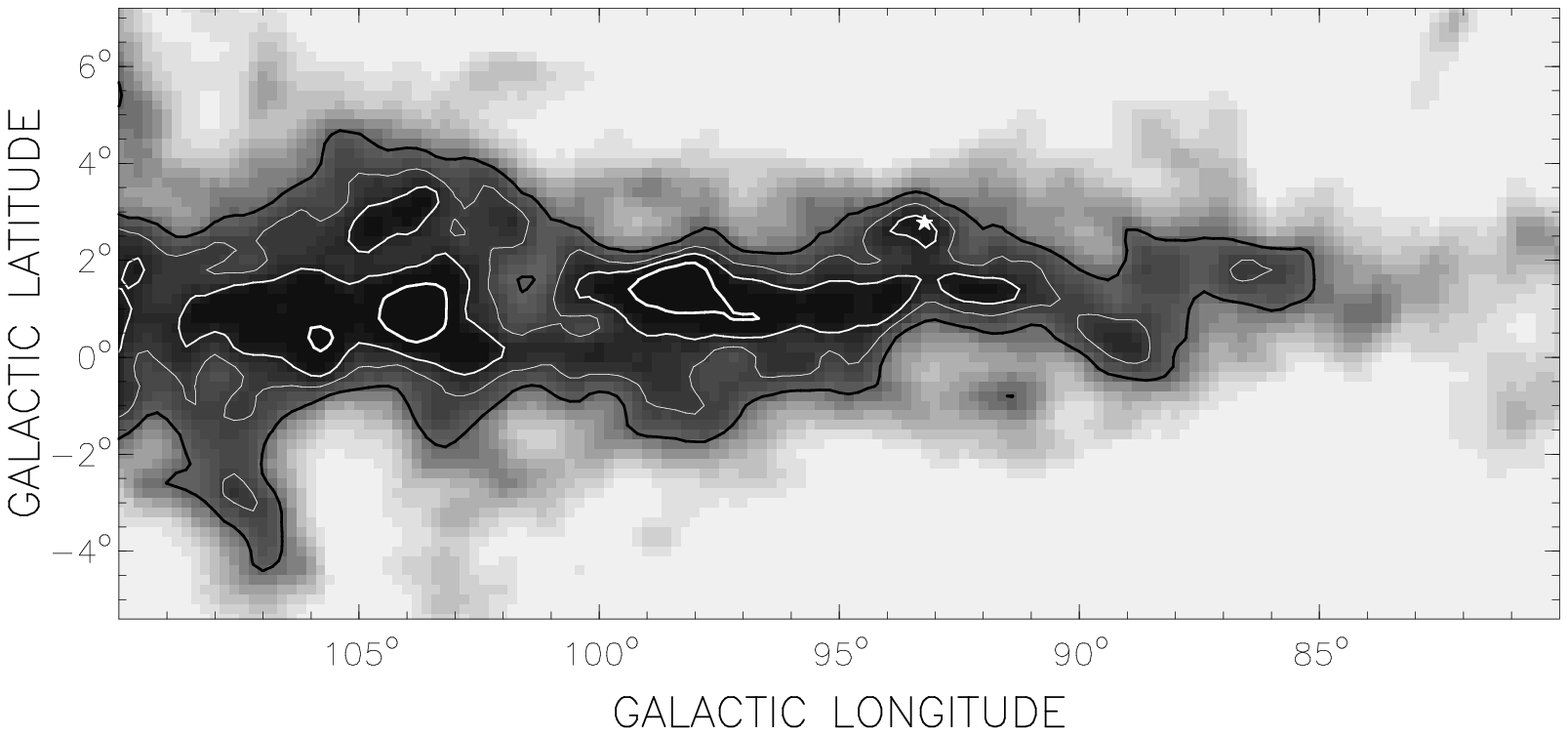}
\caption{Three views of the large-scale \ion{H}{1} environment near CTB~102,
taken from the 36$\arcmin$ resolution 26-m DRAO \ion{H}{1} survey. (Top)
$T_{B}\left(b,~V\right)$ (latitude-velocity) centered on $\ell = 93 \fdg 6$,
(Center) $T_{B}\left(\ell,~V\right)$ (longitude-velocity) centered on
$b = +2 \fdg 8$, (Bottom) $T_{B}\left(\ell,~b\right)$ (longitude-latitude)
centered at $V_{LSR}=-$58~km~s$^{-1}$. The position $\ell,b,V$ of CTB~102 is
marked with the star symbol in each slice. Brightness-temperature contours in each panel are at 40, 54, 65 and 80 K. CTB~102 appears associated with
\ion{H}{1} that could either be a high-latitude extension of the
Perseus spiral arm (centered at $-$44~km~s$^{-1}$) pushed to negative
velocities by the {}``rolling'' motion in the arm, or part of a low-velocity
extension of \ion{H}{1} from the Cygnus spiral arm ($-$68~km~s$^{-1}$),
either of which could be part of a very large \ion{H}{1} bubble expanding
away from its parent arm.}
\label{lbvhi}
\end{figure}

To begin with, Figure \ref{lbvhi} shows a latitude-velocity slice in the
direction $\ell \sim 93\fdg 6$, taken from the 26-meter telescope survey of
Galactic plane \ion{H}{1} \citep{higg00}. Extensive intermediate
velocity ($-90 \leq V_{LSR} \leq -20$~km~s$^{-1}$) \ion{H}{1} is seen
extending from $-2 \degr \leq b \leq +5 \degr$ in latitude.  The \ion{H}{1}
beyond $V_{LSR} < - 20$~km~s$^{-1}$ appears split into three blended
concentrations. The lower-latitude concentration at $b = 0 \degr$ and
$V_{LSR}= - 44$~km~s$^{-1}$ blended together with the concentration near
$b = +1 \fdg 4$ at $V_{LSR} = - 68$~km~s$^{-1}$ are contemporarily taken together
to form the Perseus \ion{H}{1} arm \citep[see][for example]{robe72}. The Outer
arm is the smaller oblong \ion{H}{1} feature up at $b= + 3 \fdg6$,
$V_{LSR}=-82$~km~s$^{-1}$, which is also blended with the upper-latitude
portion of the Perseus arm.

As seen in the top panel of Figure~\ref{lbvhi}, the contemporary Perseus arm
exhibits a very substantial tilt in latitude towards negative velocities. This
is known as the {}``rolling'' motion in this arm. As well, the splitting of the
\ion{H}{1} Perseus arm into two concentrations has been explained by a spiral
shock \citep{robe72} which is thought to precede the arm. However, an
alternative explanation for the $\ell,b,V$ structure in the \ion{H}{1} suggests
that there are actually three spiral arms beyond $-25$~km~s$^{-1}$
\citep[e.g.][]{vall08, kimes89}; the Perseus arm (centered at
$b,~V_{LSR}=0\degr,~-44$~km~s$^{-1}$), the Cygnus (=Outer) arm
($+1\fdg4,-68$~km~s$^{-1}$), and the {}``Far'' Outer arm up at $b=+3\fdg6$.
In this 3-arm interpretation, the Perseus and Cygnus arms have more reasonable
moderate rolling gradients ($dV/db \sim -2$~km~s$^{-1}$ per degree). The Cygnus
arm appears at higher latitudes than Perseus since it lies further into the
upwardly-warped outer Galactic disk.

The systemic velocity estimated for CTB~102 in \ion{H}{1}
($\simeq - 58$~km~s$^{-1}$; see Sec \ref{disc}) and its latitude ($b=+2\fdg8$)
place it somewhat in the middle of the contemporary Perseus arm, in an
\ion{H}{1} {}``wall'' that connects the two brighter concentrations.
In $b,~V$ plots (Figure~\ref{lbvhi}, top), CTB~102 is situated between the
two major \ion{H}{1} concentrations, and in the 3-arm picture above it is not
clear in which spiral arm it resides. In $\ell,~b$ maps (Figure~\ref{lbvhi}, bottom) a
{}``mushroom-cap''-like feature centered on CTB~102's coordinates
($\ell=93.2\degr$, $b=+2 \fdg 8$, $V_{LSR}=-58$~km~s$^{-1}$) appears as a
{}``blister'' off of the top of the extensive \ion{H}{1} arm at $b=+2\degr$.
As well, in an $\ell,~V$ plot of $80 \degr \leq \ell \leq 110 \degr$ (Figure
\ref{lbvhi}, center) this cap is seen as a finger-like extension off of the
Cygnus \ion{H}{1} arm, centered at $V_{LSR}=-57$~km~s$^{-1}$ extending down to
$-47$~km~s$^{-1}$. It has the appearance of either an upper-latitude extension
to the Perseus arm (up to $b=+3\degr$), or a blended \ion{H}{1} concentration
(possibly the wall of an expanding bubble) extending off the Cygnus arm to more
positive velocities. A distance to the Perseus arm in this direction then forms
a good estimate for at least a lower-limit distance to CTB~102.

The velocity of CTB~102 ($V_{LSR}\simeq-58$~km~s$^{-1}$) indicates a rather
large kinematic distance of 6.8~kpc, assuming a flat rotation curve with
$R_{0},~\Omega_{0} \simeq 7.6$~kpc, 28~km~s$^{-1}$~kpc$^{-1}$ \citep[e.g. see][]
{koth07}. However, this assumes that the systemic $V_{LSR}$ derives only from
the projection onto the line-of-sight of the object's Galactic circular motion.
Being near a major spiral feature of the Galaxy, CTB~102 is likely influenced by
gravitational forces from the stellar component of the arm, so its velocity is
likely tainted with non-circular motions such as those from the {}``rolling''
motions in the arms (velocity gradients perpendicular to the plane), and
streaming motions from the gravitational influence of density waves.

\begin{figure}
\epsscale{0.5}
\plotone{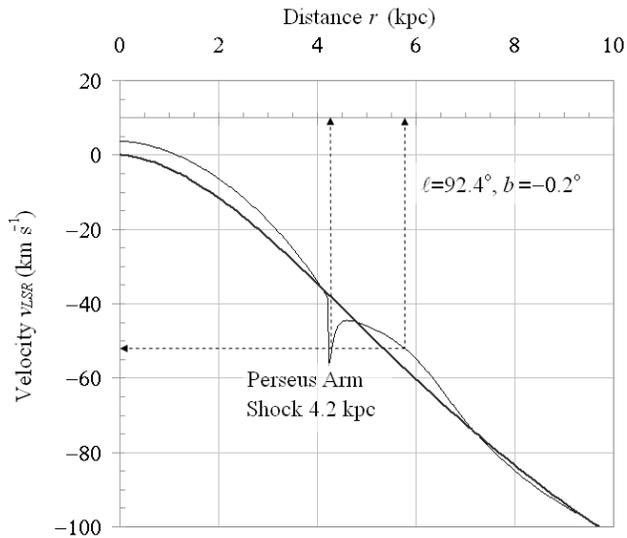}
\caption{Model velocity field towards $\ell = 92 \fdg 4,~b = -0 \fdg 2$,
from the best-fitting model of \ion{H}{1} density and kinematics of
\citet{fmac06}. The circular velocity field only (no streaming motions from density
waves) is shown by the thick line; with streaming added, the result is the thin
line which includes the shock at 4.2~kpc. Based on CTB~102's velocity
\citep[corrected for the {}``rolling'' gradient in the Perseus arm, as
in][]{fost09} of $V_{LSR}^{corr} = -$52~km~s$^{-1}$ is found at either 4.3~kpc or
5.8~kpc along the line-of-sight (assuming it is a Perseus arm object; see
text).}
\label{vfield}
\end{figure}

A new kinematic-based distance method that accounts for non-circular streaming
motions due to a two-armed density wave has been developed by \citet{fmac06}.
The approach is to model the Galactic \ion{H}{1} distribution and rotation
curve by fitting an empirical model of Galactic structure and density-wave
motions to the observations, rather than just assuming a purely circular model.
The model assumes only two arms are present in the second quadrant, arranged
as a two-armed density wave pattern. We model a direction near to $b=0\degr$
to ensure that velocity gradients with latitude \citep[i.e. rolling
motions, which are not accounted for in][]{fmac06} are not included. The
{}``rolling'' gradient of the \ion{H}{1} Perseus arm observed towards CTB~102
is $dV/db \sim -2$~km~s$^{-1}$ per degree of latitude; corrected for this
latitude gradient, CTB~102's systemic velocity is
$V_{LSR}^{corr}\simeq -52$~km~s$^{-1}$. The velocity field of the best fitting
\ion{H}{1} model in the direction $\ell = 92.4\degr$, $b= -0.2\degr$ is shown
in Figure~\ref{vfield}. In this direction only the Perseus arm is seen along
the line-of-sight. The model shows the Perseus Arm gaseous density peak
(associated with the spiral shock) to be 4.3~kpc distant from the Sun in this
direction. The velocity field is ambiguous between
$-55 \leq V_{LSR} \leq -44$~km~s$^{-1}$ due to the shock, so the heliocentric
distance to CTB~102 ($V_{LSR}^{corr}=-$52~km~s$^{-1}$) is either 4.3~kpc or
5.8~kpc.

A simple line of reasoning will show that $r= 5.8$~kpc is probably not the true
distance to CTB~102. Supposing that it is, then for $\ell=93\degr$ the
galactocentric distance to CTB~102 is $R=10.1$~kpc (for $R_{0}=8$~kpc), and in
this direction the Perseus arm shock (which defines where the arm is) is
9.3~kpc from the center. By assuming the arm is a logarithmic spiral with
pitch angle $12\degr$, one finds the intersection of a 10.1~kpc radius circle
with the shock at a current position of $\phi \approx 5 \degr$ (here $\phi$ is
galactocentric azimuth, defined as zero from the Galactic center to the Sun and
related to longitude $\ell$ by $\sin \phi = \left(r/R\right) \sin \ell$;
hence $\phi$ is positive in the 2$^{\textrm{\footnotesize{nd}}}$ quadrant of
longitude). Thus the stars have since migrated some $\Delta \phi \sim 30 \degr$ beyond the arm (their formation place) to their current position
$R,~\phi = 10.1$~kpc,~$35\degr$. How long has it taken them to move this
$30\degr$ angular distance? For {}``flat'' circular rotation the angular
velocity of CTB~102 is $\Omega = 220/10.1 = 21.8$~km~s$^{-1}$~kpc$^{-1}$. A reasonable pattern
speed for the Perseus arm is $\Omega_{p}=15$~km~s$^{-1}$~kpc$^{-1}$, a
mean between modern estimates near 20 \citep[e.g.][]{mart04} and older
estimates near 11 \citep[e.g.][]{robe72} for the Milky Way's spiral pattern. Then the
time it took the system to migrate $\Delta \phi$ from the arm at angular rate
$\Omega-\Omega_{p}$ (the relative angular velocity with respect to the arm) is
$t=\Delta\phi/\left(\Omega-\Omega_{p}\right)=75$~Myr. If the massive stars in
CTB~102 formed from the compression of the shock, then by now the only ones
left would be those that live longer than 75~Myr, or stars of
$\lesssim 7$~M$_{\odot}$ (B5V types, for example). A cluster of hundreds of such stars would be needed to
power CTB~102 ($U = 112$~pc~cm$^{-2}$; see Sec. \ref{central}). It is highly unlikely that CTB~102 contains stars later
than B5 \textit{only}. It is more likely that CTB~102 lies near to the
arm/shock, which is 4.3~kpc distant in this direction.

The uncertainty in this estimate is $\pm20\%$, from the variation of best-fitting
models of the same direction and in models fitted to several immediately
adjacent directions \citep[see][]{fmac06}. This agrees with stellar distances
of the nearby optically-brilliant \ion{H}{2} regions Sh2-124
($\ell,~b=94 \fdg 6$, $-1 \fdg 5$, $V_{CO}=-44$~km~s$^{-1}$,
$r = 3.6 \pm 0.6$~kpc) and Sh2-132 ($\ell,~b = 103 \degr$, $-0 \fdg 8$,
$V_{CO} = -49$~km~s$^{-1}$, $r = 3.5 \pm 0.9$~kpc), both of which are Perseus arm
objects \citep[$^{12}$CO/\ion{H}{1}-based velocities and spectrophotometric
distances to 100 Sharpless objects in the second quadrant are forthcoming
in][]{fost09}.

We note that CTB~102 and filaments may instead be kinematically associated with
the Cygnus spiral arm, participating in an extensive star-formation group
along with \ion{H}{2} region NRAO~655
\citep[$V_{LSR} \sim -70$~km~s$^{-1}$;][]{fost01} and SNR 3C434.1
\citep[$V_{LSR} \simeq -80$~km~s$^{-1}$;][]{fost04}. This would extend its
distance estimate here by a factor of 1.6, since the mean stellar distance to
Cygnus arm \ion{H}{2} regions between $90 \degr \leq \ell \leq 100 \degr$ (Sh2-121,
BFS~8, Sh2-128, BFS~10 \& DA~568) is $r \simeq 7.0$~kpc.

\section{Radio Continuum Analysis}\label{estimates}
Radio continuum measurements of \ion{H}{2} regions can provide a lower limit on the total number of ionizing photons ($N_L$) and estimates of the average excitation parameter ($U$). Assuming an optically thin, spherical, constant-density \ion{H}{2} region, the equations used are \citep{rud96}:
\begin{eqnarray}
N_L &=& 7.5 \times 10^{43} S_{\nu} d^2 \nu^{0.1} T_e^{-0.45} ~\textrm{s}^{-1}\\
U &=& 1.33 S_{\nu}^{1/3} d^{2/3} \nu^{1/30} T_e^{0.116} ~\textrm{pc} ~ \textrm{cm}^{-2}
\end{eqnarray}
where $\nu$ is the frequency in GHz, $S_{\nu}$ is the flux density measured at frequency $\nu$ in mJy, $d$ is the distance to the source in kpc, and $T_e$ is the electron temperature in units of $10^4$ K. Parameter values used are $\nu = 1.42$~GHz, $d = 4.3$~kpc and $T_e = 10000$~K. Since none of the parameters are very sensitive to $T_e$, and the uncertainties in Table~\ref{der-tbl} are large, a general value of $10^4$ K is used \citep{rud96}.

The flux density for the considered region is obtained from the CGPS continuum image by using the Dominion Radio Astrophysical Observatory (DRAO) software \textsc{imview}. A polygon incorporating the region of interest is drawn by eye. The polygon's perimeter defines the background, and a twisted plane is fitted to estimate the background. The flux density with the estimated background subtracted is then used to compute the parameters $N_L$ and $U$. The biggest contribution to the uncertainty is where to draw the polygon around the region of interest. To estimate this uncertainty, each flux measurement is done ten times, each time with a new polygon drawn. The average value is used to compute the estimated parameters, and the standard deviation is used to estimate the uncertainty in the flux measurement.

\subsection{CTB~102 Central Region}\label{central}
\begin{figure}
\plotone{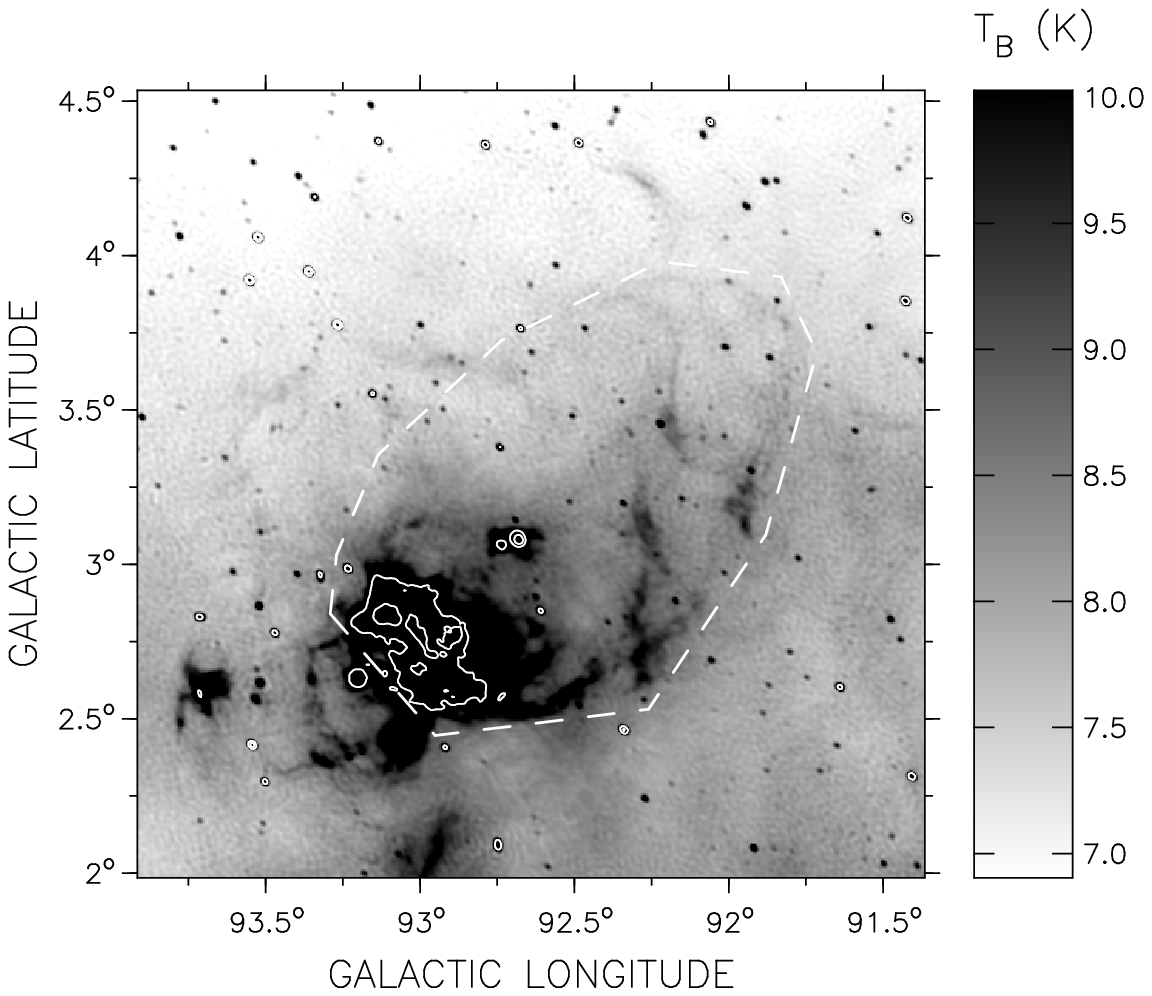}
\plotone{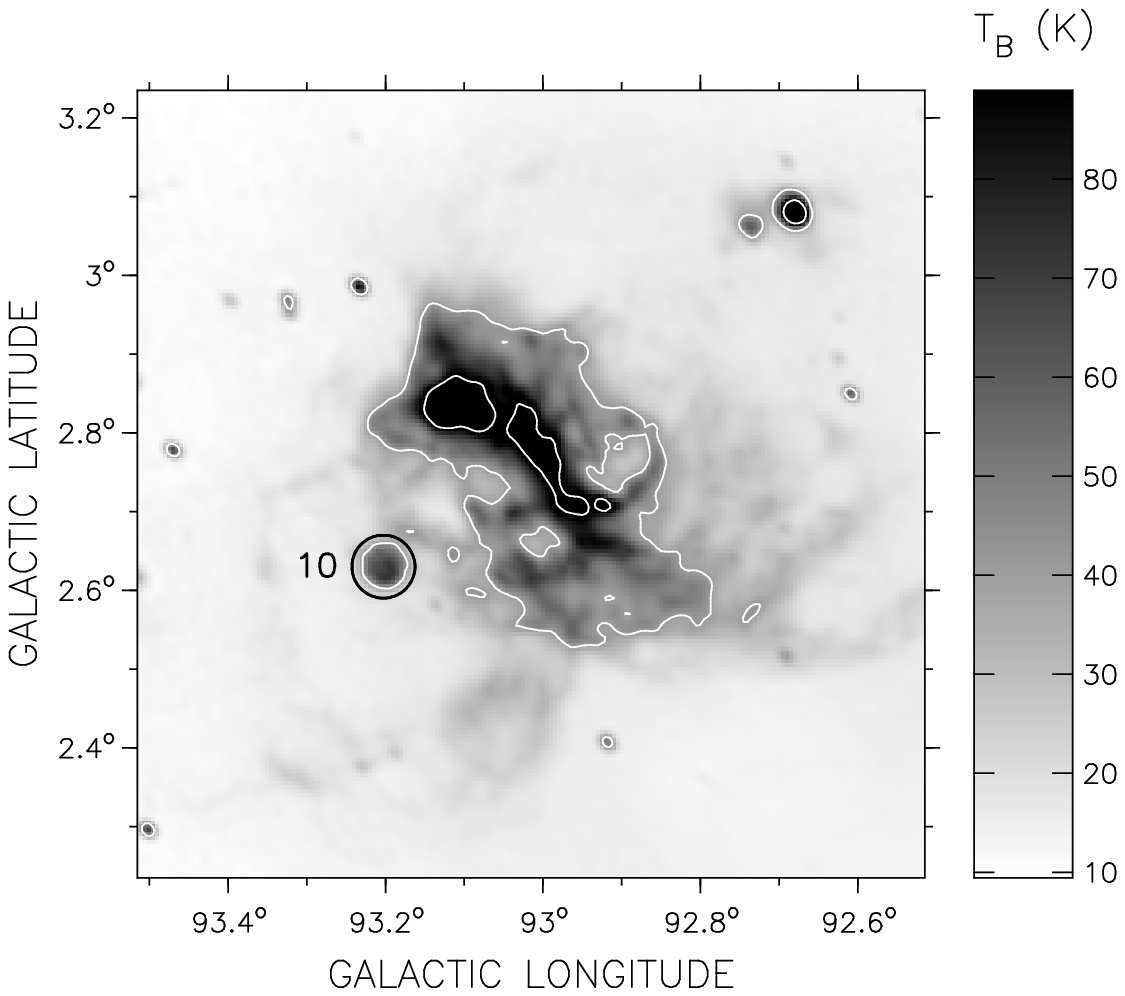}
\caption{Two CGPS 1.42 GHz continuum images of CTB~102 overlayed by contours corresponding to 15 and 30 K. (Top) The dashed polygon is a representative of the polygons used in Section~\ref{central}. (Bottom) The circle indicates the area used for estimates for filament 10 in Section~\ref{fil10}.\label{est-plot}}
\end{figure}

A representative polygon around the region of interest is shown in Figure~\ref{est-plot}, top. The flux density is measured ten times with results ranging between $28.9$ and $36.9$ Jy. After subtraction of the flux density from the foreground source WB~43, the average value and $1 \sigma$ uncertainty in the flux density is $31.3 \pm 2.5$ Jy, which gives a value of $N_L = \left( 4.5 \pm 1.8 \right) \times 10^{49}$~s$^{-1}$ and $U = 112 \pm 15$~pc cm$^{-2}$. Almost all the uncertainty in these values comes from the uncertainty in the distance. Comparing the values $ \left( \log{N_L} = 49.7 \right)$ to \citet{pa73} suggests an ionizing flux and excitation parameter corresponding to a O5 V class star. Comparing to \citet{vgs96} suggests an O4 V star. However, the 30 K contour in Figure \ref{est-plot} shows the brightest parts of CTB~102 being along an arc stretching from CTB102p towards filament 7 (i.e. towards smaller Galactic longitude and latitude). Given this continuum flux distribution, it is more likely that the ionizing flux comes from several stars. In summary, the estimates together with the morphology of the 1.42 GHz continuum emission, suggest that the CTB~102 complex is powered by at least several late type O stars, maybe even an early type O star (since $N_L$ is a lower limit).

\subsection{Filament 10}\label{fil10}
Filament 10 looks strikingly circular in the 1.42 GHz continuum image (see Figure~\ref{est-plot}, bottom) with an angular diameter of $\sim 5 \arcmin$, yielding a physical size of $\sim 6$ pc at a distance of $4.3$ kpc. The intensity profile, with a rapid fall-off at the edge, strongly suggests the region is ionization bound. As such this filament would not contribute ionizing photons to the larger complex. Drawing a circular polygon around Filament 10 (average background $T_b \simeq 11.1$ K) yields $0.46 \pm 0.03$ Jy for the flux density (same method for uncertainty estimate as for the whole complex), giving $N_L = \left( 6.6 \pm 2.6 \right) \times 10^{47}$~s$^{-1}$ $\left( \log{ N_L } = 47.8 \right)$ and $U = 28 \pm 4$~pc cm$^{-2}$. Comparisons suggest an O9.5 V \citep{pa73} or an B0.5 V \citep{vgs96} star ionizing the region. For this filament, a single star or a compact cluster of early type B stars are ionizing the probably homogeneous ISM, given the circular radio continuum profile.

\section{Discussion}\label{disc}
This study was partly motivated by the appearance of the CTB~102 region in the 1.42 GHz continuum mosaic (Figure~\ref{pointings}). Looking at Figure~\ref{velocities}, one can make out the region looking like a cone-shaped bubble. The radio brightness is concentrated along a ridge following an arc with a maximum at CTB102p. The arc goes towards filaments (in turn) 7, 6, 5, 4 and continues towards greater Galactic latitude, with faint filamentary structures around $b = +4 \degr$ eventually bending back towards greater Galactic longitude and then down. Within the arc, there appears to be a ``cleared out'' region extending from the central region of CTB~102 towards greater Galactic latitude and smaller Galactic longitude. The observed velocities for the filaments along this arc all fall within $\la 11$ km s$^{-1}$ of the velocity of CTB102p, indicating that they at least are in the same part of the Galaxy. Taking the velocities and 1.42 GHz continuum image together, it is likely that the CTB~102 complex includes filaments 7, 6, 5 and 4. Filament 2 is probably not part of the region, since its velocity deviates by $16$ km s$^{-1}$, and no obvious association with the other filaments is seen in the radio continuum data. As will be discussed below, our model shows that filament 2 could not be powered by CTB~102.

With the inclusion of these filaments, the angular size of the CTB~102 \ion{H}{2} region is $\sim 80 \arcmin$, which corresponds to a physical size of $\sim 100$ pc at a distance of 4.3 kpc. Inclusion of filaments 8, 9, and 10, all within $\la 6$ km s$^{-1}$ of CTB102p with apparent connections to CTB~102 in the 1.42 GHz image, increases the angular size of the \ion{H}{2} to $\sim 100 \arcmin$ or $\sim 130$ pc at 4.3 kpc.  The objects KR~4, KR~6 and NRAO~652 have no obvious connection to the region in the radio data, and are probably \ion{H}{2} regions which happen to be in the same part of the Galaxy as CTB~102.

If the assumption of the filaments being part of the structure is indeed correct, what sort of process would create such a big \ion{H}{2} region? The typical sound speed in a \ion{H}{2} region is on the order of $10$ km s$^{-1}$, which is also the typical velocity associated with blister regions and champagne flows \citep{tenorio82}. This velocity corresponds to $\sim10$ pc/($10^6$ yr). The velocity is consistent with the observed velocities in Table~\ref{par-tbl}. However, to reach a radius of $50-65$ pc ($2R= \sim 100$ or $\sim 130$ pc) would require a time scale of 5 to 6.5 million years (assuming constant expansion velocity). This time scale is longer than the main-sequence life time of a O5 V  star, $3.7 \times 10^6$ years \citep{schaller92}, but comparing CGPS 408 MHz and 1.42 GHz show no sign of a supernova in the CTB~102 region. Also, CTB~102 does not appear like a blister in the 1.42 GHz continuum image.

Rather, an alternative interpretation is the whole CTB~102 complex with its filaments is a combination of an \ion{H}{2} region with a bubble/chimney structure. A stellar wind from massive a O star(s) creates a bubble in the interstellar medium (ISM). Towards a region of the Galaxy with lower density (higher Galactic latitude) the expanding bubble has its top ``blown off'', creating the possible chimney. Towards regions with higher densities, the expanding stellar wind bubble sweep up a shell of the ISM. The shell and left over neutral material is exposed to ultraviolet radiation from the O star(s), gets ionized and forms the \ion{H}{2} region.

What would be expected if this interpretation is correct? The \ion{H}{1} structure would show a cavity in \ion{H}{1}. In this picture, the expanding bubble sweeps up a shell of neutral material. This shell would be ionized from the inside of the bubble and if viewed through the edge, the increased path length would cause the shell to appear as neutral and ionized filaments. These \ion{H}{1} and \ion{H}{2} filaments would have observed velocities which do not deviate too much from each other and from the velocity of the central region. Also, the ionized filaments would be expected to be located on the ``inside'' of the neutral filaments.

\begin{figure}
\plotone{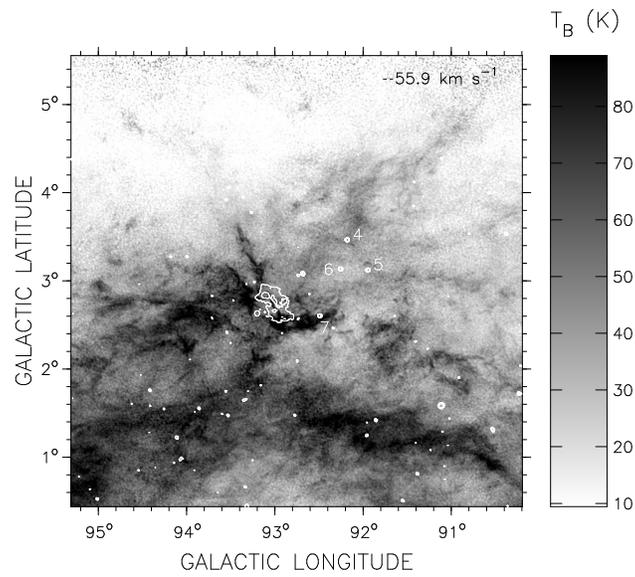}
\caption{CGPS \ion{H}{1} line image, averaged over three velocity channels to a width of 2.5~km~s$^{-1}$, centered on the velocity shown in the frame. The white contours correspond to 1.42 GHz continuum brightness temperature levels of 15 and 30 K. The white circles indicate the positions of the GBT pointings for filaments 4, 5, 6 and 7 (from Figure~\ref{pointings}). This figure is reduced due to size constraints. For the full version, see the published article in The Astrophysical Journal.\label{h1-figure}}
\end{figure}

Support for the idea comes from viewing the \ion{H}{1} structure in the region. Figure~\ref{h1-figure} shows the averaged velocity slices (of width 2.5 km s$^{-1}$) of the CGPS \ion{H}{1} line data, overlayed with contours (15 and 30 K) indicating where the 1.42 GHz continuum flux is located in relation to the \ion{H}{1}. A cleared out region is most easily seen in the final three velocity frames ($-60.8$ to
$-65.8$ km s$^{-1}$), extending from the central region of CTB~102 towards the Galactic north. This cavity in \ion{H}{1} disappears beyond $\sim -72$ km~s$^{-1}$. The Galaxy contains gas out to $\sim -100$ km~s$^{-1}$ in this direction (see Figure~\ref{lbvhi}, center). The \ion{H}{1} emission is mostly concentrated along an arc extending from north of CTB~102, wrapping around the ``edges'' of the \ion{H}{2} region and extending more faintly up next to filaments 7, 6 and 4. These \ion{H}{1} filaments (seen in the final three velocity frames, $-60.8$ to
$-65.8$ km s$^{-1}$) are located on the ``outside'' of the corresponding \ion{H}{2} filaments and they correspond roughly in velocity to the RRL results in Table~\ref{par-tbl}. In the first velocity frame, $-53.4$ km s$^{-1}$, a filament of \ion{H}{1} is spatially coincident with filament 5 (observed $V_{LSR}=-51.9$ km~s$^{-1}$).

\begin{figure}
\plotone{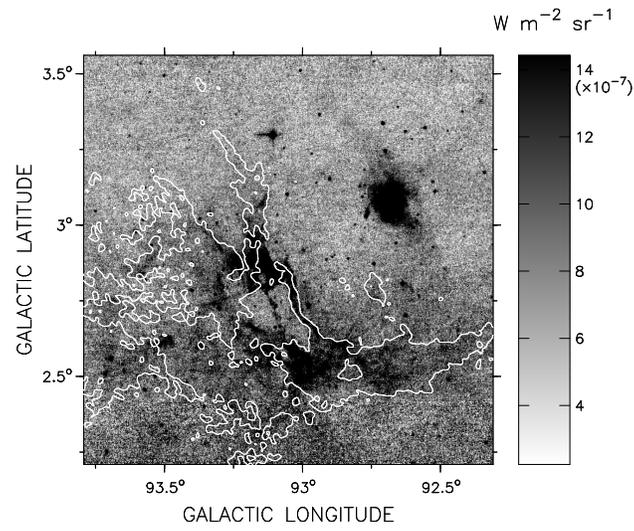}
\caption{MSX A-band image (8.3 \micron). The contour corresponds to an averaged CGPS \ion{H}{1} brightness temperature level of 70 K from the $-55.9$ km s$^{-1}$ velocity frame in Figure~\ref{h1-figure}.\label{msx}}
\end{figure}

Figure~\ref{msx} shows a MSX A-band image (8.3 \micron) towards CTB~102. The contour corresponds to an averaged CGPS \ion{H}{1} brightness temperature level of 70 K from the $-55.9$ km s$^{-1}$ velocity frame in Figure~\ref{h1-figure}. The MSX A-band traces PAH emission from neutral material being bombarded with ultraviolet photons. Figure~\ref{msx} clearly shows the morphological association between structures seen in the \ion{H}{1} arc discussed above, neutral material in a UV radiation field and the 1.42 GHz continuum emission. This is best demonstrated by the 8.3 \micron~emission tracing a thin filamentary \ion{H}{1} structure which extends from the main \ion{H}{1} arc towards CTB102p (i.e. towards greater Galactic longitude and latitude). This thin filament runs parallel to the bright arc seen in the 1.42 GHz continuum image (traced by the 30 K contour in Figure~\ref{est-plot}), something which would be expected if the interpretation of a neutral shell being ionized from slightly above the bright arc is correct.

Can ionizing photons reach the filaments? To reach the inside of the proposed swept up shell at filament 6, ionizing photons have to travel an angular distance of $\sim 50 \arcmin$ from CTB102p, corresponding to $\sim 60$ pc at $4.3$ kpc. In the bubble scenario, the inside of the bubble is filled with stellar wind, a hot, low density plasma. Absorption in such a plasma is negligible, and the purely geometrically diluted radiation field could easily maintain a high ionization fraction in the ionized inside of the swept up shell. The same is true for the other filaments we consider part of the shell structure. For filament 2, at a distance of $\sim 180$ pc from CTB102p ($\sim 140 \arcmin$), a purely geometrically diluted radiation field could maintain a high ionization fraction if the stellar wind bubble extended all the way out to filament 2. The proposed bubble does not extend that far (see below), and inserting any realistic ISM between the proposed bubble and filament 2 causes significant absorption in the ISM, thus leaving no photons from CTB~102 to ionize filament 2.

A simple analytical model of the development of a stellar wind bubble was given by \citet{cmw75}. Using the formulation in \citet{kwok}, the radius ($R$), shell expansion velocity ($\dot{R}$) and shell fractional thickness ($ \Delta R / R$) of such a region are given by:
\begin{eqnarray}
R &=& 28 \left( \frac{\dot{E} / 10^{36}~ \textrm{erg~s}^{-1}}{\mu n_0 /1~ \textrm{cm}^{-3}}   \right)^{\frac{1}{5}} \left(  t/10^6 ~\textrm{yr}\right)^{\frac{3}{5}} ~\textrm{pc}\\
\dot{R} &=& 16 \left( \frac{\dot{E} / 10^{36}~ \textrm{erg~s}^{-1}}{\mu n_0 /1~ \textrm{cm}^{-3}}   \right)^{\frac{1}{5}} \left(  t/10^6 ~\textrm{yr}\right)^{- \frac{2}{5}} ~\textrm{km} ~\textrm{s}^{-1}\\
\frac{\Delta R}{R} &=& \frac{1}{3} \left( \frac{a_0}{\dot{R}} \right)^2
\end{eqnarray}
where $\dot{E}$ is the energy loss rate from the star, $\mu$ is the mean molecular weight of the medium the shell is expanding into ($\mu = 1.4$ for neutral ISM, $\mu = 0.62$ for a fully ionized ISM), $n_0$ is the ISM density and $a_0$ is the sound speed inside the bubble.

We estimate $\dot{E}$ using a typical mass loss rate of $\dot{M} = 10^{-6}$ M$_{\sun}$/yr and a stellar wind velocity of $V = 2000$ km s$^{-1}$. The sound speed is $a_0 = 10$ km s$^{-1}$ and the ISM density is $n_0 = 1$ cm$^{-3}$ \citep[mean ISM density, from][]{basu99}. If $\mu = 1.4$ it would take $3.8 \times 10^6$ years to reach the position of filament 6 from CTB102p (60 pc), by that time the expansion velocity of the shell will be $9$ km s$^{-1}$. The other extreme, if $\mu = 0.62$, a radius of 60 pc would be reached in $2.8 \times 10^6$ years, with a expansion velocity at that time of $12$ km s$^{-1}$. The expansion velocity for both scenarios is fairly insensitive to large $t$, and reaches order of magnitude $\sim 10$ km s$^{-1}$ after roughly 2.5 million years. The estimated thickness of the proposed shell from the 1.42 GHz continuum image and the \ion{H}{1} line data is $\sim 6 \arcmin$, corresponding to a fractional thickness of $\sim 12 \%$. The time it takes the model shell to reach such a fractional thickness is $0.9 \times 10^6$ years (if $\mu = 1.4$) or $1.3 \times 10^6$ years (if $\mu = 0.62$). If instead $n_0 = 10$ cm$^{-3}$ (the rest of the parameters remain the same), time to reach 60 pc increases by a factor of $\sim 4$. This suggests that if this mechanism is responsible for the observed filaments, the ISM in the direction of decreasing Galactic latitude cannot be more than a few particles per cubic centimeter, or the timescale exceeds the main-sequence life time of the probable star(s).

In the basic model for the expanding swept-up shell, the shell density would be \citep{kwok}:
\begin{equation}
\rho_s = \left( \frac{\dot{R}}{a_0} \right)^2 \rho_0 ~ \textrm{cm}^{-3}
\end{equation}
Estimates of the column density test this picture where the observed filaments are due to our line of sight through the edge of the shell. The maximum path length $s$ looking through the edge of a uniform density spherical shell of radius $R$ and thickness $\Delta R$ would be:
\begin{equation}
s = 2 R \sqrt{2 \left( \Delta R / R \right) + \left( \Delta R / R \right)^2 }
\end{equation}
Estimating the column density towards the filaments is done by integrating the CGPS \ion{H}{1} line data. The single channel (width 0.82 km s$^{-1}$) rms noise is 3.3 K, corresponding to a single channel rms noise in the column density of $4.9 \times 10^{18}$ cm$^{-2}$. For filament 6, $R = 60$ pc and $\Delta R / R = 0.12$, which gives $s \approx 60$ pc. This is on the order of the radius of the proposed bubble in this direction, so this is probably an upper limit in path length through the edge of any shell associated with CTB~102. The column density through filament 6 is $\sim 10^{21}$ cm$^{-2}$ (making the rms noise negligible). The uniform density needed in a shell to produce such a column density is $\sim 5$ cm$^{-3}$. In the model, the time required to reach this shell density (with $n_0 = 1$ cm$^{-3}$) is $0.6 \times 10^6$ years. Given the large uncertainties of the parameters involved, in $R$ (from the distance uncertainty), $\Delta R$ (uncertainty in telling where the filaments start and end) and column density (mainly determining the background), the uncertainty in the shell density can be as large as $\sim 60 \%$. Within such a large uncertainty, the model can account for observed column densities within a time frame set by the main-sequence life time of a massive star. However, it cannot be ruled out that the observed filaments are higher density regions in the ISM that happen to be exposed to ionizing photons from the star(s) powering the \ion{H}{2} region. Even if that is the case, the medium between the filaments and the powering source would most likely be low density material with a high ionization fraction to allow for enough photons to reach the filaments. Such a medium is consistent with the \ion{H}{1} line data and the 1.42 GHz continuum image.

The \citet{cmw75} model is very basic and does not take into account multiple stellar winds nor non-uniform ISM, but the estimates indicate that the bubble model is a plausible explanation of the size of CTB~102, allowing for observed size, velocities and fractional shell thickness in a time scale of $1-2 \times 10^6$ years. That is less than the main-sequence life time of the star(s) that is probably powering the region. For the same reason, it is unlikely that the proposed bubble would reach all the way out to filament 2. The overall morphology of CTB~102 indicates that the proposed \ion{H}{2} region/bubble/chimney is a major feature in the Perseus Galactic arm, powerful enough to disrupt the ISM and clear out a $\sim 100-130$ pc region. The appearance of CTB~102 in the \ion{H}{1} with a cleared out region makes the CTB~102 complex comparable in size to the W4 superbubble in the Perseus arm \citep{norman96}. The W4 superbubble is modeled by \citet{basu99} as expanding in a stratified atmosphere. An association of nine O-type stars (IC 1805) is probably the reason for the cavity in \ion{H}{1} and the ionization structure of the W4 \ion{H}{2} region. CTB~102 shares similarities with W4 in large scale \ion{H}{1} distribution and ionization structure.

\section{Summary}
We have obtained new RRL observations of CTB~102 that show that the filamentary structure surrounding the central region is physically associated with the central region. We provide the first ever distance estimate for this \ion{H}{2} region, 4.3 kpc. We argue that the best explanation for the size and appearance of the whole complex is that it is a large \ion{H}{2} region combined with a wind-blown interstellar bubble/chimeny structure. MSX, \ion{H}{1} and 1.42 GHz observational data are consistent with this view.

\begin{acknowledgements}
The authors thank Dana Balser of NRAO, Green Bank,
for his help and problem solving skills during the GBT observations. We thank the referee for many helpful comments. The National Radio Astronomy Observatory is a facility of the National Science Foundation operated under cooperative agreement by Associated Universities, Inc.
The Dominion Radio Astrophysical Observatory is operated as a national facility
by the National Research Council of Canada. The Canadian Galactic Plane Survey
is a Canadian project with international partners, and is supported by a grant
from the Natural Sciences and Engineering Research Council of Canada (NSERC).
This work made use of data products from the Midcourse Space Experiment.
\end{acknowledgements}

\clearpage

\clearpage

\clearpage

\clearpage

\clearpage

\clearpage

\clearpage

\clearpage

\clearpage

\begin{deluxetable}{ccccccc}
\tablecolumns{7}
\tabletypesize{\footnotesize}
\tablewidth{0pc}
\tablecaption{Observations.\label{obs-tbl}}
\tablehead{
\colhead{Filament/} & \colhead{RA} & \colhead{Dec}   & \colhead{$\ell$}    & \colhead{$b$}    &\colhead{Integration time} & \colhead{$\overline{T}_{system}$}\\
\colhead{Object} & \colhead{h m s} & \colhead{$\degr$ $\arcmin$ $\arcsec$}   & \colhead{$\degr$}    & \colhead{$\degr$} & \colhead{s} & \colhead{K}}
\startdata
CTB102p&
21 12 28.9&
52 32 23&
93.115&
2.835&
3600&
23.7\\
2&
21 01 32.5&
50 51 17&
90.725&
2.955&
30000&
21.9\\
4&
21 05 19.1&
52 17 01&
92.185&
3.465&
14400&
21.3\\
5&
21 05 56.4&
51 52 53&
91.950&
3.125&
15000&
21.0\\
6&
21 07 14.6&
52 07 03&
92.260&
3.135&
10200&
21.1\\
7&
21 10 48.6&
51 55 51&
92.495&
2.605&
3000&
21.9\\
8&
21 15 39.9&
52 22 03&
93.325&
2.365&
7200&
22.4\\
9&
21 13 59.6&
52 15 09&
93.065&
2.470&
2400&
23.0\\
10&
21 13 49.2&
52 27 25&
93.195&
2.630&
600&
23.5\\
KR~4&
21 16 24.3&
52 48 02&
93.715&
2.585&
3000&
22.1\\
KR~6&
21 19 01.0&
53 22 26&
94.400&
2.705&
6600&
21.5\\
NRAO~652&
21 15 20.1&
51 52 56&
92.940&
2.065&
6600&
23.5\\
WB~43&
21 09 21.6&
52 22 22&
92.668&
3.069&
600&
23.9\\
\enddata
\tablecomments{Positions are J2000. Integration times and average system temperatures are for the GBT observations.}
\end{deluxetable}

\clearpage

\begin{deluxetable}{ccccccc}
\tablecolumns{7}
\tabletypesize{\footnotesize}
\tablewidth{0pc}
\tablecaption{Observed line parameters.\label{par-tbl}}
\tablehead{
\colhead{Filament/} & \colhead{$T_l$} & \colhead{$\Delta T_{rms}$}   & \colhead{$V_{LSR}$}    & \colhead{$\Delta V$}    &\colhead{$|V-V_{ref}|$}\\
\colhead{Object} & \colhead{(mK)} & \colhead{(mK)}   & \colhead{(km~s$^{-1}$)}    & \colhead{(km~s$^{-1}$)} & \colhead{(km~s$^{-1}$)}}
\startdata
CTB102p&
$102.6 \pm 0.4$&
$1.2$&
$-62.66 \pm 0.05$&
$18.34 \pm 0.08$&
\nodata\\
2&
$1.7 \pm 0.1$&
$0.3$&
$-78.4 \pm 0.6$&
$13.9 \pm 1.1$&
16\\
4&
$2.6 \pm 0.1$&
$0.6$&
$-67.2 \pm 1.3$&
$30.6 \pm 2.2$&
5\\
5&
$3.5 \pm 0.2$&
$0.7$&
$-51.9 \pm 0.9$&
$21.0 \pm 1.5$&
11\\
6&
$3.5 \pm 0.2$&
$0.6$&
$-67.0 \pm 0.8$&
$25.4 \pm 1.4$&
4\\
7&
$6.7 \pm 0.3$&
$1.2$&
$-57.7 \pm 0.8$&
$22.8 \pm 1.4$&
5\\
8&
$4.4 \pm 0.2$&
$0.7$&
$-56.4 \pm 0.7$&
$21.2 \pm 1.2$&
6\\
9&
$10.7 \pm 0.3$&
$1.1$&
$-66.3 \pm 0.5$&
$24.8 \pm 0.8$&
4\\
10&
$29.0 \pm 0.5$&
$1.5$&
$-65.0 \pm 0.2$&
$19.7 \pm 0.4$&
2\\
KR~4&
$12.1 \pm 0.3$&
$0.9$&
$-67.8 \pm 0.4$&
$24.5 \pm 0.6$&
5\\
KR~6&
$4.7 \pm 0.2$&
$0.7$&
$-56.1 \pm 0.5$&
$16.0 \pm 0.9$&
7\\
NRAO~652&
$5.7 \pm 0.2$&
$0.7$&
$-68.7 \pm 0.5$&
$26.0 \pm 0.9$&
6\\
WB~43&
$68.9 \pm 0.6$&
$1.7$&
$-0.8 \pm 0.1$&
$15.3 \pm 0.2$&
62\\
\enddata
\tablecomments{Spectral parameters for each filament/object observed. Line amplitude ($T_l$), central velocity ($V_{LSR}$) and FWHM ($\Delta V$) are obtained by a Gaussian fit to the radio recombination line after baseline subtraction, regridding and smoothing. The noise, $\Delta T_{rms}$, is obtained by considering regions on both sides of the line. Both $T_l$ and $\Delta T_{rms}$ are given in brightness temperature units. The uncertainties correspond to $1 \sigma$. The $|V-V_{ref}|$ column displays the absolute difference in $V_{LSR}$ for each observed filament/object, compared to CTB102p and rounded to the nearest integer.}
\end{deluxetable}

\clearpage

\begin{deluxetable}{cccc}
\tablecolumns{4}
\tabletypesize{\footnotesize}
\tablewidth{0pc}
\tablecaption{Derived parameters.\label{der-tbl}}
\tablehead{
\colhead{Filament/} & \colhead{$T_D$} & \colhead{$T_e$}  \\
\colhead{Object} & \colhead{(K)} & \colhead{(K)}}
\startdata
CTB102p&
$7400 \pm 100$&
$5200_{-200}^{+200}$
\\
2&
$4200 \pm 600$&
$3400_{-3400}^{+12200}$
\\
4&
$21000 \pm 2900$&
$5300_{-4100}^{+3600}$
\\
5&
$9800 \pm 1300$&
$6000_{-4300}^{+3800}$
\\
6&
$14400 \pm 1500$&
$5000_{-3600}^{+3200}$
\\
7&
$11600 \pm 1300$&
$5000_{-2000}^{+1900}$
\\
8&
$10000 \pm 1100$&
$8300_{-3100}^{+2900}$
\\
9&
$13800 \pm 800$&
$7900_{-1100}^{+1000}$
\\
10&
$8600 \pm 300$&
$7700_{-500}^{+500}$
\\
KR~4&
$13400 \pm 600$&
$8100_{-1000}^{+900}$
\\
KR~6&
$5600 \pm 600$&
$5800_{-4300}^{+3800}$
\\
NRAO~652&
$15100 \pm 1000$&
$6000_{-2000}^{+1900}$
\\
WB~43&
$5200 \pm 100$&
$5500_{-300}^{+300}$
\\
\enddata
\tablecomments{Derived parameters for each filament observed. The uncertainties in $T_D$ are $1 \sigma$ uncertainties. The listed uncertainties in $T_e$ comes from considering uncertainty in the background estimate, 0.5 K. The total uncertainty in $T_e$ are by far dominated by this background uncertainty, the other two contributions (from $T_l$ and $\Delta V$) are both $\simeq 10 \%$ of the dominant background contribution.}
\end{deluxetable}

\end{document}